\documentclass{cambridge6A}
\usepackage{amsmath,amssymb}
\usepackage{graphicx}
\usepackage{booktabs}
\usepackage{hyperref}
\usepackage{textcomp}

\def\etal{{\sl et al.}}
\newcommand{\met}{\mbox{\ensuremath{\slash\kern-.7emE_{T}}}}

\begin{document}
\setcounter{chapter}{13}
\begin{center}
{\Large\textbf{The Search for the Top quark}}\\
\vspace{5mm}
  \textbf{Paul D. Grannis} \\ 
  \textit{Stony Brook University}
\end{center}


The discovery of the top quark in 1995 completed the roster of the matter particles,  quarks and leptons,
expected in the Standard Model and in many of its suggested extensions.  
However, the top quark does not
 conform to the pattern set by its sister quarks and has come to be a critical component of the ongoing
 searches for physics beyond the prevailing paradigm.

\noindent
({\it Prepared for the Proceedings of the 4th International Symposium on the History of Particle Physics, CERN, Nov. 2025.})

\begin{center}
{\bf Prologue}
\end{center}

The  current model  of  hadrons~\cite{symposium3}  began coming  into focus in early 1964 when Murray 
Gell-Mann~\cite{gellmann1964}, and George Zweig~\cite{zweig1964} independently posited three 
fractionally-charged constituents of the known mesons and baryons~\cite{llewellyn-smith}.
They hypothesized that these hadrons filled multiplets based on the representations 
of SU(3) and that the quarks themselves (or `aces' in  Zweig's terminology) 
were contained in the fundamental representation.  
They noted that the octets of pseudoscalar and vector mesons, as well as the 
spin 1/2 baryons, could be constructed by various combinations of the 
up ($u$), down ($d$) and strange ($s$) quarks whose 
charges were $(+2/3)e$, $(-1/3)e$  and $(-1/3)e$ respectively.

Gell-Mann did not give his quarks a physical reality.  
``It is fun to speculate about the way quarks would behave if they were physical particles (instead of purely
 mathematical utilities)"  he wrote. 
``A search for stable quarks of charge $-1/3$ or $+2/3$ at the highest energy accelerators would help 
to reassure us of the non-existence of real quarks. ”~\cite{gellmann1964}.  
Zweig gave his aces a more concrete role in building hadrons, writing  
``On the other hand, saying that the vectors or aces are some kind of spurions, 
fictional particles that help in computing consequences of symmetry, is also not correct.  
Aces, unlike conventional spurions, bind and have physically observable mass differences."~\cite{zweig1964}
But early experiments searching for fractionally charged particles gave null results, so most
physicists accepted the view that the quarks were mathematical devices, useful for explaining
the patterns of the hadrons.

Zweig's picture of quarks as concrete physical entities was however dramatically 
confirmed by the MIT-SLAC deep inelastic
$ep$ scattering experiments of the late 1960's and early 1970's 
led by Jerome Friedman, Henry Kendall and Richard Taylor.
They observed excess scattering at large momentum transfers which, 
similarly to the Rutherford experiment, was attributed to point-like entities within the proton.   
Subsequent studies demonstrated that these entities were spin $1/2$ and fractionally charged, in 
agreement with the Gell-Mann/Zweig prescription.    
The advent of quantum chromodynamics, the theory governing the strong interaction
with its property of color confinement, explained the puzzle that the quarks 
had not been seen as free objects in nature.  
The prevailing view of quarks as physical constituents of matter confined within hadrons 
now became widely accepted and
Gell-Mann qualified his view :
``I employed the term ‘mathematical’ for quarks that would not emerge singly and ‘real’ 
for quarks that would."~\cite{gell-mann1992}

In 1964, James Bjorken and Sheldon Glashow had already recognized~\cite{bjorken-glashow} 
that there was a missing element in the quark picture.  They proposed a new `charm' quark, $c$,  
so that the number of quarks would be the same as the number of leptons.
The necessity of charm was underscored in 1970 by 
Glashow, John Iliopoulos and Luciano Maiani who showed that a $c$-quark isospin partner for the $s$-quark would explain 
the observed lack  of strangeness-changing neutral current reactions.   
A more general argument constraining the pattern of quarks 
arises from  the theoretical observation that the renormalizability of the underlying
field theory requires a symmetric structure of lepton doublets 
and the three color copies of quark doublets.

The tension was dramatically relieved in the 1974 ‘November revolution’ discoveries of a neutral
3.1 GeV meson, both in $pp$ collisions by Samuel C.C. Ting's group at Brookhaven Lab~\cite{J_discovery} 
and in $e^+e^-$ collisions by Burton Richter's group at SLAC ~\cite{psi_discovery},  called respectively
$J$ and $\psi$.    
Its appearance in $s$-channel $e^+e^-$ collisions identified it as a $J^P=1^-$ state like the 
$\rho, \omega$ and $\phi$ mesons, 
but with a width about 1000 times too small to be an ordinary strong decay.  
The $J/\psi$ was identified  as the `charmonium’ cousin of positronium  consisting of a new charge $(+2/3)e$ 
charm quark and its antiquark, each with a mass of about half that of the $J/\psi$.   The existence
of charm was solidified with the 1976 discovery of the open-charm $D$ mesons.

The closure afforded by the two symmetric lepton and quark generations was short-lived.  In 1976, 
Martin Perl and collaborators at SLAC discovered~\cite{tau_discovery} 
the tau lepton in $e^+e^-$ collision events with a final state of an 
electron, muon  and missing energy arising from the decays 
of pair-produced tau leptons such as $\tau^+ \rightarrow \mu^+ \nu_\mu \bar{\nu}_\tau$ and   
$\tau^- \rightarrow e^- \bar{\nu}_e \nu_\tau$ .   
This new asymmetry between lepton and quark generations 
was exacerbated with the 1976 discovery at Fermilab
by Leon Lederman and collaborators~\cite{upsilon_discovery}  of a 10 GeV narrow resonance in $p$ Be interactions 
in the final state $\mu^+ \mu^-$.  Similarly to charmonium, this Upsilon state was quickly understood to be the 
bound state of a new charge $(-1/3)e$ $b$ (bottom) quark and its antiquark.  

\begin{center}
{\bf Early Top Quark Searches}
\end{center}

This new gap in the table of fermions was obvious and it
did not require a rocket scientist to predict the existence of a partner to the bottom quark
and to name it top.     But as there was no theory to predict the fermion masses, one resorted 
to guesses rooted in previous experience.  Since the Upsilon had about three times the 
mass of the $J/\psi$, which in turn is about three time heavier than the $\phi$ 
(the $s \bar s$ quarkonium state), it was generally expected that the toponium state 
formed from $t$ and $\bar t$ would be about three times the Upsilon mass
and the top quark mass would be roughly half of that, or about 15 GeV.

Searches in $e^+e^-$ collisions began in earnest~\cite{quadt}. 
The experiments at DESY's PETRA collider set a lower limit of $m_t>23.3$ GeV.    
The TRISTAN collider was built at KEK with its main purpose to
find the top quark, but settled for a limit of 30.2 GeV.    The LEP and SLC experiments observed no 
decays of the $Z$  boson into top pairs and  raised the top mass limit to  45.8 GeV.    
     
With no further $e^+e^-$  colliders in view, the searches turned to the 
higher energy hadron colliders now coming into operation.   
In 1976, Carlo Rubbia, Peter McIntyre and David Cline~\cite{ppbar-collider} 
suggested the possibility of creating antiprotons from collisions of protons in a 
heavy metal target, collecting and cooling them either by co-moving beams of electrons 
as proposed by Gersh Budker~\cite{budker} or using the stochastic method of 
Simon van der Meer~\cite{vandermeer}, 
and introducing them in counter-rotating directions into the existing proton 
accelerator to provide high energy $p\bar p$ collisions.  
CERN embraced the proposal and in 1978 set about building the antiproton facility to 
operate in the SPS using stochastic cooling for the antiprotons.   Operation of the S$p$$\bar p$S  
began in 1981 at center-of-mass energy $\sqrt s = 546$ GeV.                   

There were two general purpose experiments at the S$p$$\bar p$S.  
The larger UA1 led by Rubbia employed tracking detectors within a 0.7 T dipole magnet, four separate calorimeters, and muon detectors installed in the iron yoke magnetic field return.  UA1 pioneered 
the concept of hermeticity as a means of sensing the presence of non-interacting particles 
through missing transverse energy, $\met$~\cite{met}.  
The smaller and complementary UA2 experiment led by Pierre Darriulat emphasized state-of-the-art 
segmented EM and hadronic calorimetry.  
There was no central magnetic field in UA2, but there were air core toroids in the forward and backward regions.~\cite{jenni}. 

In 1984, UA1 showed `clear evidence’ ~\cite{ua1-1984} for an enhancement in 
$p\bar p \rightarrow \ell \nu +2$ jets which was consistent with the decay
$W \rightarrow t \bar b$  for $m_t$ in the range $30 - 50$ GeV.   
The jets were the collimated sprays of particles resulting from quarks that 
have neutralized their color as they evolve into hadrons.
This excess was subsequently attributed to underestimated backgrounds~\cite{ua1-1988},
 and a limit $m_t > 44$ GeV was set.    
In 1990, with the  S$p$$\bar p$S now operating at $\sqrt s =630$ GeV, 
UA1~\cite{ua1-1990} and UA2~\cite{ua2-1990} 
 obtained improved limits of 60 and 69 GeV respectively, again searching for $W \rightarrow t b$.

\begin{center}
{\bf Fermilab Top Quark Searches}
\end{center}

 There had been discussions at Fermilab during the 1970’s on the possibility of a collider, with many
options considered:  colliding protons with protons, antiprotons or electrons, and the use of
the existing conventional-magnet Main Ring or  a new superconducting magnet ring.  
But at this time the Laboratory was still focused on bringing the original fixed target program up to speed 
so  it was judged to be premature to embark on the addition of a collider ~\cite{hoddeson}.   
Seeing the rapid progress toward  the CERN S$p$$\bar p$S, Fermilab subsequently opted for a 
higher energy collider in a superconducting magnet ring, housed in the existing tunnel below
the Main Ring, thus ceding the initial top quark searches (as well as the discoveries of the $W$ and $Z$ bosons)
to CERN.  The Tevatron was approved in 1982, saw first collisions in 1985 and began physics operation 
in 1988 with the low luminosity Run 0 at $\sqrt s = 1.8$ TeV,  followed by a longer Run 1 beginning in 1992.   
The later addition of the Main Injector to replace the old Main Ring and various other improvements were completed for Run 2 that began operations at $\sqrt s = 1.96$ TeV in 2001~\cite{holmes}. 

Initially only one general purpose detector was foreseen at the Tevatron.   In 1976 Director Wilson
established a Coliding Beams Department within the Research Division to develop a detector
proposal for the Main Ring version~\cite{hoddeson}.  However the Accelerator Division had the responsibility for
the superconducting magnet rings and associated experimental infrastructure.  The two organizations had
different visions of the nature of an appropriate future collider 
and this impaired progress for a couple of years.
By 1978, now with Lederman as director, the choice of  protons and antiprotons circulating in
a new superconducting magnet ring was made,    With the  collider design fixed, the  
Collider Detector Facility, led by Alvin Tollestrup and Roy Schwitters was able to converge 
on a detector design and submit the CDF detector design report in 1981~\cite{cdf_design_staley}.
The collaboration included several prominent US labs and universities and groups from Japan and Italy.    
These strong groups retained much of the power to decide their own contributions so that 
CDF developed a relatively loose central authority in its decision making~\cite{hoddeson}.  
The CDF design included a 1.6 T solenoid magnet outside a small-cell drift chamber, 
electromagnetic and hadronic calorimeters based on scintillator or proportional chamber
readout with lead or iron absorbers, 
and muon detectors in the large polar angle regions.   
For Run 1, an innovative silicon microstrip vertex detector, SVX,  was added, enabling the 
identification of short-lived hadrons containing a bottom quark.   

In 1981, after a visit to CERN and the two experiments at the S$p$$\bar p$S, Lederman called for a 
`small ($7\times7\times10$ m$^3$), simple and clever’ experiment that would
run for a year or two at the D0 interaction point 
where the extraction magnets for fixed target beams were located.   
In 1983 all 19 proposals were rejected by the PAC and I was appointed to form 
a collaboration to develop a more ambitious experiment.   
The nascent collaboration failed to coalesce on a name
and settled for its address in the Tevatron lattice.   
With its larger set of smaller collaborating groups, D0 developed a more top-down 
management structure~\cite{hoddeson}.  With the collider fully specified, the D0 detector
design was able to proceed more straightforwardly than that of CDF.  However
CDF's use of much of the available technical manpower and funding caused
severe strains on the D0 progress.

The 1984 D0 design report described a set of tracking chambers and a 
transition radiation detector with no surrounding magnet, a finely segmented, hermetic, uranium-liquid argon
sampling calorimeter and large solid-angle magnetized iron toroids for muon detection.  
The design was neither small nor simple, and clearly anticipated more than a two year lifetime!  
 There were insufficient funds to raise the Main Ring above D0 as had been done for CDF
so the 400 GeV protons threaded their way through the calorimeters,  and data collection 
had to be disabled during their traversals.  Although D0 was only commissioned for the 
1992 Run I, four years after CDF began, the subsequent rapid increase in luminosity allowed the two experiments 
to compete on an equal footing.

It is interesting to look back to the CDF and D0 design reports to see what they said about the top quark. 
 CDF’s report wrote that a $t\bar t$ toponium resonance would be observable if its mass were less than 30 GeV
 (but noted that such a mass was already excluded by PETRA results).  
It discussed the observation of top-flavored mesons with mass 30 GeV, 
and also noted that quarks heavier than the $W$ boson would decay preferentially 
to a $W$ and a lighter quark but did not offer any limits on the observable mass reach.  
The D0 report discussed the case of $W \rightarrow t \bar b$ for $m_t=30$ and 60 GeV, and claimed 
that pair production of heavy fourth generation quarks with $m_Q=120$ GeV and decays 
to $W$ plus a light quark could be observed (but only in the experimentally challenging case 
that both $W$’s decay hadronically).   Both collaborations seem to have been convinced 
that the top would be discovered by the S$p$$\bar p$S experiments before the 
Tevatron program got underway.  And both significantly underestimated the integrated luminosity 
that the Tevatron would deliver.  Although the study of the $W$ and $Z$ bosons and high transverse
momentum jets were the 
primary targets used for designing the experiments, both detectors turned out to be very well suited to find
and study the top quark.

By 1990 the action turned to the Tevatron, with its three-fold energy advantage over CERN.
Using Run 0 data, CDF increased the mass limit in  1991 to 77 GeV~\cite{cdf-1991}, 
thus ruling out the possibility that  $W \rightarrow t \bar b$.   
The searches now shifted to  focus on the strong interaction production of top-antitop pairs.   
With the increased luminosity in Tevatron Run 1, CDF reported a 91 GeV lower limit~\cite{cdf-1992} followed by
the last-ever limit of 131 GeV by D0~\cite{d0-1994}.

In parallel with the direct searches for the top quark, predictions for its mass were made in the 
context of the Standard Model, using precision observables such as the $Z$ boson properties 
measured at LEP and SLC~\cite{pepe}, 
as well as deep-inelastic $e N$ and $\nu N$ scattering, $\nu e$ scattering 
and atomic parity violation interactions~\cite{quadt}.   
These measurements over-constrained the electroweak sector of the Standard Model and provided rather tight predictions 
for possible top quark masses, generally just above the contemporaneous experimental limits~\cite{quigg}.
 In March 1994, on the eve of the first direct evidence for the top being presented, the indirect estimate 
was $172^{+22}_{-24}$ GeV.

Since the CKM matrix elements coupling top to the down or strange quarks are very small,
a high mass top quark almost always decays to a $W$ boson and a $b$ quark, and thus the final states in 
$t\bar t$ production are governed solely by the way the two $W$ bosons decay.  
The $W$ boson decays equally to all available fermion-antifermion pairs, so the decays to the three lepton pairs
 ($e\nu_e$, $\mu\nu_\mu$, $\tau\nu_\tau$)  are half as frequent as those to the six quark pairs 
(three color combinations of $ud_w$ and $cs_w$, where the subscript $w$ denotes the 
quark weak eigenstates) with each channel occuring with a probability of 1/9.  
The dilepton final states (excluding those involving the more difficult to identify $\tau$) account for 5\%
of $t\bar t$ events and the $e+$jet and $\mu$+jet final states comprise 30\%.    The dilepton channel
has relatively little background. The lepton+jet channels have appreciable background,
largely from QCD production of $W$+jets; but with just one unseen final state particle 
kinematic reconstruction of the $t\bar t$ state is possible.
The 44\% of the final states with no leptons suffer from huge QCD multijet backgrounds.  These
non-leptonic final states were ultimately observed but have played
little role in precision top quark studies.
   
By 1992, a few tantalizing events with top-quark-like features 
and very low background probabilitiers began to surface.  
 CDF recorded an event  with an electron, four jets and \met ~in which two of the jets 
were tagged as $b$-jets by displaced vertices seen in the SVX.
Assuming this to be $t\bar t$ production and decay, the computed top quark mass was $175\pm10$  GeV.  
D0 reported a striking event with a 100 GeV electron, a 200 GeV muon, 100 GeV \met~and two jets.  
Assuming a $t\bar t$ origin, the top mass was estimated to lie in the range $145 - 200$ GeV.

\begin{figure}
\includegraphics[width=0.7\textwidth]{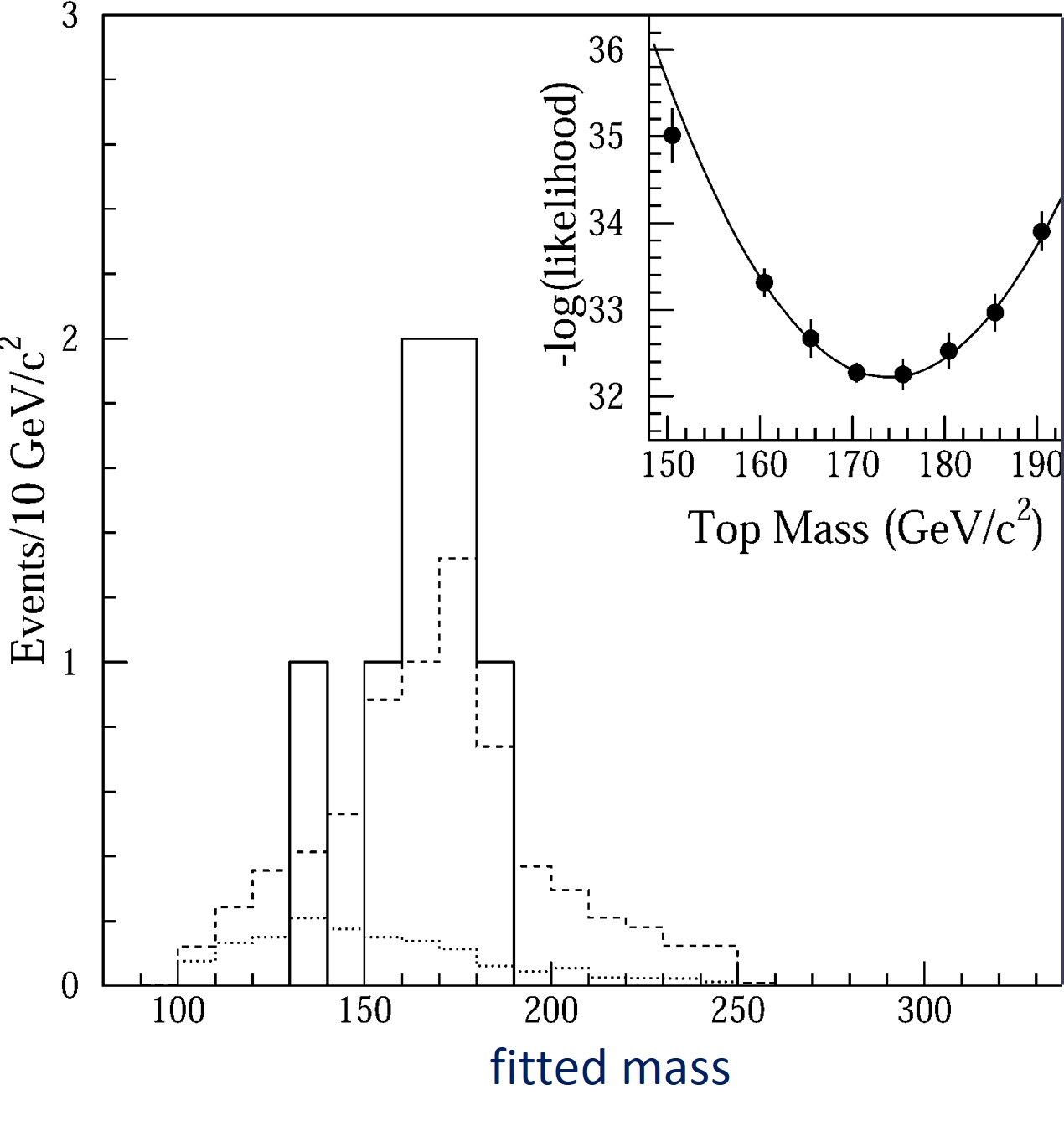}
\caption{ 
 Fitted mass distribution of $b$-tagged lepton+jets top quark candidates 
in the 1994 CDF `Evidence' paper (solid line).  
The predicted background and top quark signal distributions (dotted lines) are shown.  
The inset shows the log-likelihood distribution versus the hypothesized mass.
(Reprinted with permission from Ref.~\cite{CDF-evidence_PRL}, copyright 1994 by 
the American Physical Society.)
\label{CDF-fittedmass} 
}
\end{figure}

In August 1993, three CDF  analysis subgroups brought  internal reports 
to the full collaboration, now led by William Carithers and Melvyn Shochet, 
showing small event excesses over background predictions.  
For the dilepton channels, two events were seen with 0.56 events predicted.   
For lepton+jets channels, two categories of events containing evidence of a $b$-jet were defined.   
The SVX tag selected events containing a separated vertex consistent with $B$ meson decay. 
There were six SVX tags observed, with a background of 2.3. 
The SLT tag selected events in which a low $p_T$ lepton was seen close to a jet, 
consistent with semileptonic decay of a $B$ hadron.  
Seven SLT tags were observed while only 3.1 were expected from background sources.  
Three events had both SVX and SLT tags.   
In the lepton plus jets sample, the hypothesized $t\bar t$ pairs were reconstructed
from the observed lepton, jets and \met, and the top quark mass was obtained
from a two-constraint fit.
The distribution of the fitted masses for observed lepton+jet events is shown in 
Fig.~\ref{CDF-fittedmass} and compared with the expectations of background 
and the top pair production hypotheses.
Many in CDF were skeptical and numerous cross-checks were performed to buttress the analyses.  
After eight months of internal deliberations and extensive discussions of the words to be used 
in the title (`Search’, `Evidence’, `Observation’  ...) the paper was submitted in May 1994~\cite{cdf_evidence_staley}.    
The  signal significance was 2.8$\sigma$ and the top quark mass was measured to be $174\pm 16$ GeV,
with a cross section somewhat higher than expected~\cite{CDF-evidence_PRL,CDF-evidence_PRD}.

D0 reported its findings at the August 1994 ICHEP Conference in Glasgow~\cite{grannis-ichep}. 
With no central magnetic field and no displaced vertex $b$-quark tags, 
D0 relied on the topological variables aplanarity ($\cal A$) and $H_T$ to suppress the backgrounds 
in their lepton+jets analysis.  
$\cal A$ is proportional to smallest eigenvalue of the momentum tensor of the jets in an event
and is expected to be large for the relatively isotropic $t\bar t$ event decay products 
and small for the more collinear $W$+jets background.   
$H_T$, defined as the scalar sum of the jet $E_T$'s, is larger for the decay of the 
high mass top decay products than for the backgrounds in which the jets 
are typically due to gluon radiation.  
The D0 sample had a comparable expected significance to that of CDF 
but the data showed only a  1.5$\sigma$ excess over the background expectations.

\begin{center}
{\bf The Discovery of the Top Quark}
\end{center}

During the shutdown at the end of summer 1994, the Fermilab Accelerator Division 
discovered that one of the Tevatron magnets in the ring had been rotated around the beam axis.   
Fixing this immediately boosted the luminosity delivered to the experiments and by the end of the year, 
the accumulated data sets had approximately tripled.  
When the data were examined for conferences in January 1995, 
CDF and D0 realized that a discovery was now eithin reach 
and feverish efforts to complete the analyses began.

A year or so earlier, a procedure had been put in place by John Peoples, now the Fermilab Director,
that if one collaboration brought a discovery paper,
 a one week period would begin for the other collaboration 
to finish its paper and submit for simultaneous publication.   
This arrangement removed much of the fear of being scooped, and helped
ensure that the papers were more fully discussed and cross-checked.  
This mechanism followed the experience at the ISR in 1973 when the CERN-Rome group 
waited for a couple of weeks for the competing Pisa-Stony Brook group to complete 
its analysis showing the total $pp$ cross section to rise with energy at high energies.

Nevertheless, considerable anxiety remained.   
CDF, sensing that the junior D0 experiment was catching up, was determined that
the long gestation of its evidence paper not be repeated.
D0 displayed its paranoia by hiding its analysis files in a QCD group directory
 PRJ\$\_ROOT227:[QCD\_3.CUCKOO].   
An internal discussion in Collaboration X about initiating desirable cross checks elicited the comment   
``It was important that we understand these differences but when it was suggested 
that more work was needed on this matter, the answer was that there wasn't time 
because [Collaboration Y] was about to go public!”
The flow of information from one collaboration to the other was minimal:  
a couple with one spouse in CDF and the other in D0 claimed there was no discussion of the 
ongoing studies at home.    There were however some probes:  
one spokesman asked his Administrative Assistant 
to phone the AA of the other group to ask if work on a paper draft was underway.

CDF retained the analysis used for the 1994 evidence paper.  
D0 continued to use $\cal A$ and $H_T$ cuts to reduce background but re-optimized them for 
the higher expected top mass.  Working independently but on parallel tracks both collaborations 
were closing in on their results by mid February.  
On Feb. 17, CDF brought a proposed paper submission to Peoples and the one-week clock began ticking.  
D0 had a draft in hand but spent a few more days doing some final cross-checks.  

On Feb. 24, CDF and D0 simultaneously submitted their papers to Physical Review Letters.   
Arranging the public announcement was complicated by the fact that 
while the papers were being submitted, a symposium on charm, bottom 
and top quark studies was in progress in Rio de Janeiro and many CDF and D0 people, 
as well as Peoples,  were there.  
Giving the talks on top were excrutiating -- speakers had to present the outmoded earlier
search results knowing that the discovery was already in the can.

\begin{figure}[h!]
\centering
\includegraphics[width=0.75\textwidth]{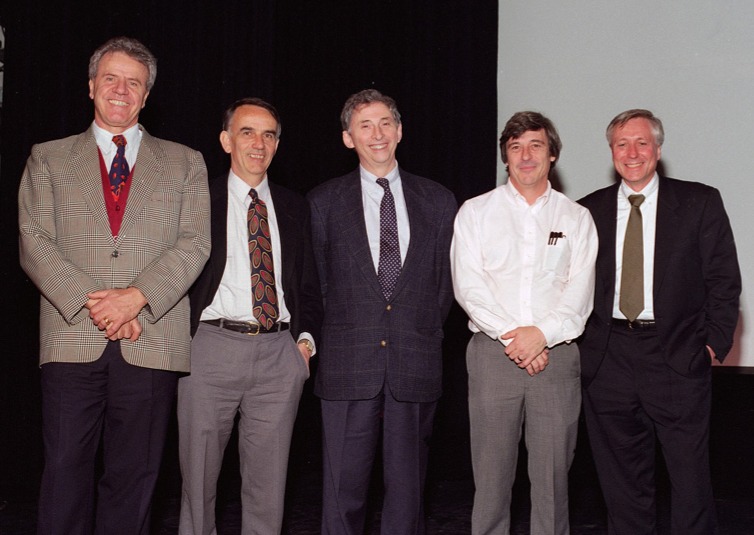}
\caption{ 
(L-R):  Giorgio Bellettini, CDF co-spokesman,
Paul Grannis, D0 co-spokesman, John Peoples, Fermilab Director, Hugh Montgomery, D0-co-spokesman,
Bill Carithers, CDF co-spokesman at the top quark discovery seminar.   Photograph by Reidar Hahn, Fermilab;
used with permission of  Fermilab Creative Services.
\label{spokes} 
}
\end{figure}

The public announcement was  scheduled 
for the following week when the principals were back in Fermilab
and results were embargoed in an attempt to keep them under wraps.   
But a reporter from the Los Angeles Times called a prominent Fermilab physicist for an 
explanation of `statistical evidence’ as it related to the O.J. Simpson trial. 
He replied using the ‘top quark discovery’ as an example and the cat was out of the bag.
Many newspapers swarmed to get the story, including a full court press by the Brazilian papers 
that wanted a scoop with a local connection.  
A major US paper threatened not to cover the eventual announcement if they did not get the story.   
On March 2, 1995 the co-spokesmen of CDF and D0 (Fig.~\ref{spokes}) made the public announcement of the  
discovery to an overflow audience in the Fermilab auditorium and received extensive press coverage,
even from outlets that had threatened to abstain.

Both discovery papers~\cite{cdf-obs,d0-obs}  featured observed fitted mass distributions similar to
that in Fig.~\ref{CDF-fittedmass}
that showed a clear excess of events over those associated with known backgrounds.   
CDF had 50 events 
(44 single lepton, all with vertex or soft muon $b$-tags and 6 dilepton) and an 
expected background of $23.4 \pm 6.4$ events. They obtained
$\sigma_{t\bar t} = 6.8^{+3.6}_{-2.4}$ pb and a top quark mass of $176 \pm 13$ GeV.  
The D0 analysis, with more restrictive selection cuts, 
had 17 events (14 single lepton with topological or soft muon $b$-tags and 3 dilepton) and an 
expected background of $3.8 \pm 0.6$ events.  They reported $\sigma_{t\bar t} =6.4 \pm 2.2$ pb and 
a top quark mass of $199^{+29}_{-30}$ GeV.  
Both papers showed fitted mass distributions with relaxed cuts that allowed more 
background but still with a clear signal.    

The significances for an excess over predicted backgrounds were 
4.8$\sigma$ and 4.6$\sigma$ for CDF and D0  respectively.   
To the best of my knowledge, the top quark significances
were the basis for the now-canonical rules of 5$\sigma$ for discovery and 3$\sigma$ for evidence – 
although neither of the 1995 top observations nor the 1994  evidence  quite 
reached those benchmarks!

\begin{figure}
\includegraphics[width=0.8\textwidth]{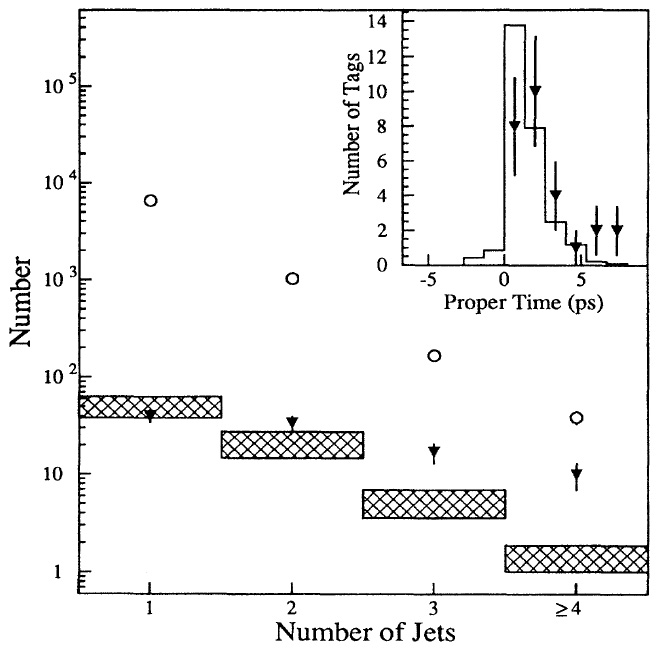}
\caption{ 
 Distribution of CDF $W$+jets events as a function of the number of jets.  
The inset shows the proper time distribution of those jets tagged a $b$-jets.
(Reprinted with permission from Ref.~\cite{cdf-obs}, copyright 1995 by 
the American Physical Society.)
\label{CDF-added} 
}
\end{figure}

\begin{figure}
\includegraphics[width=0.75\textwidth]{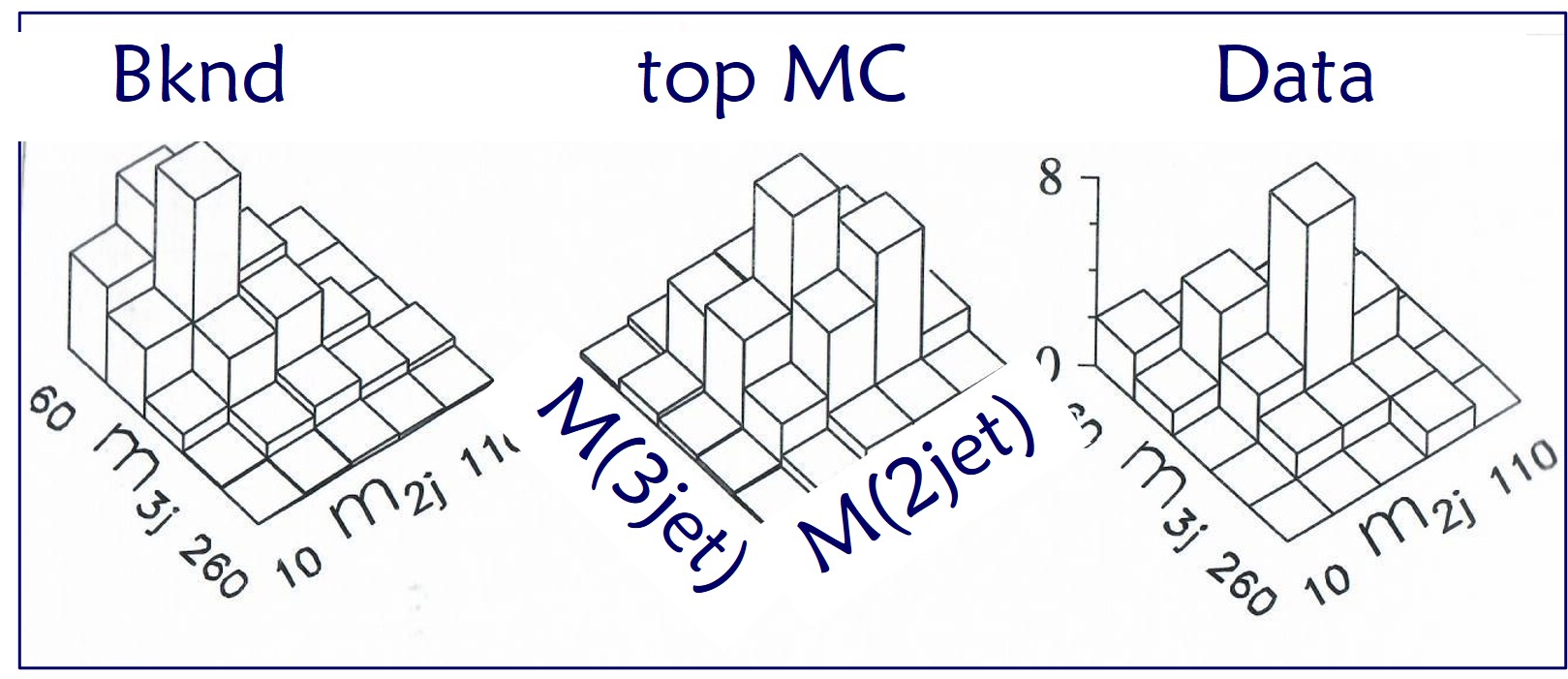}
\caption{ 
Two dimensional distribution of the hadronically decaying top candidate in
the  D0 lepton+jets events as a function of
the three jet mass and the smallest two jet mass.   The predicted
distributions for the background, top quark of mass 200 GeV and the loose sample data are shown.
(Reprinted with permission from Ref.~\cite{d0-obs}, copyright 1995 by 
the American Physical Society.)
\label{D0-added} 
}
\end{figure}

Both discovery papers offered additional evidence beyond that given by the fitted mass plots.   
The distribution of CDF  $W$+jets events as a function of the number of jets, $N_{\rm jet}$, shown
in Fig. ~\ref{CDF-added}  shows good agreement with QCD backgrounds for $N_{\rm jet} \leq 2$ 
but an excess for $N_{\rm jet} \geq 3$ where the top events would be expected.  
The inset shows the distribution of proper times for candidate $b$-quark hadron tags, 
in good agreement with the known $B$ lifetime.   
Figure~\ref{D0-added}  shows two-dimensional distributions of the D0 two-jet and three-jet masses 
for the hadronically decaying top candidate in a loose single lepton sample,
together with background and signal distributions.  
The data  clearly show the presence of a subset of events consistent with top and
its decay $W$ boson, in addition to a background component at smaller masses.

\begin{center}
{\bf Concluding Remarks}
\end{center}

In the succeeding years, refined measurements of the top quark mass, cross sections, 
branching fractions, polarization, spin correlations and decay $W$ boson helicity were 
measured by the Tevatron experiments.   Of particular note was the observation of
single top quark production~\cite{single_top_D0,single_top_CDF} 
by the electroweak interaction. 
This enabled the measurement of the top quark lifetime of
about $5\times10^{-25}$s, much shorter than the time required for it to bind with other quarks into hadrons.
Thus, even the expectation that quarks are `mathematical’ in the sense of not appearing 
free in nature needed revision.  Our understanding of quarks has morphed from 
virtual entities convenient for classification, 
to always bound entities forbidden to emerge singly,
to real objects that can exist as isolated entities even if for only a brief instant.  

The LHC experiments have now added significantly to our understanding
of the top quark and as we look forward to future energy frontier colliders, the top quark studies will 
continue to be a major element of their programs.
As always, yesterday’s discoveries become today’s calibration tools 
(and backgrounds for searches for even rarer phenomena).   

It is impossible to single out just a few key people responsible for the top discovery.  
CDF and D0 each had over 400 physicists in 1995.  
The contributions of those who designed and built the detectors, devised the systems to operate them, 
manned the control rooms 24-7 for years, wrote the software for
analyzing the data and managed the collaborations 
were also essential.  We were disappointed that a proposal to award the 
American Physical Society’s Panofsky Prize in Particle Physics to the collaborations was unsuccessful 
(awards were then limited to three or fewer people), but were gratified that the 
2019 European Physical Society gave its 2019 High Energy and Particle Physics Prize to 
the CDF and D0 Collaborations for 
``{\sl The discovery of the top quark and the detailed measurements of its properties}”.  
This award not only recognized the essential contributions 
of everyone active at the time of the discovery, but also those 
who came later and filled in the portrait of the top quark.

A fall 1995 account in the SLAC Beam Line~\cite{beamline} traced the path to discovery 
and gave a flavor of the personalities and internal workings of the collaborations.   
In a guest editorial, Bjorken offered some observations on the top discovery from his vantage point 
as a close observer (as the Fermilab Associate Director of Physics and as a participant 
in an experiment located midway between CDF and D0):

\noindent
``The history of physics is full of near-simultaneous discoveries by separate individuals or groups, and with that 
often has come acrimony and controversy, from Newton and Leibnitz to Richter and Ting, and down to the present time. 
There has been competition between CDF and D0 as well.  ...~~ [but]  the ensuing CDF/D0 competition 
has served for constructive purposes; I have never seen this competitiveness to be corrosive. ...  
This piece of competition has been a class act.  
Not only has this been true between the collaborations, but it seems also to have been the case within them.
This is no mean feat, since harmony within a big group of strong individualistic physicists of great talent 
and often even greater ego is not easy to maintain."

I think that Bjorken raises an interesting point.   It is perhaps surprising that hundreds of individuals, working
in  a self-organized group without the authority inherent in conventional institutions,  can work effectively
to produce significant scientific advances.   Certainly achieving that coherence requires effort and
the willingness to carefully consider the opinions of fellow collaborators.   But the impetus given by
the opportunity to make significant discoveries makes the collaborations powerful entities that are themselves worthy 
of study.


\begin{thebibliography}{}
\bibitem{met}
`Missing transverse energy' is defined as~  $\met = -\Sigma_i p_{T,i}^2$, where 
$p_{T,i}$ is the momentum perpendicular to the initial beams' directions of the $i_{\rm th}$
final state object.   
The name is a misnomer since energy is a scalar and
can have no association with only transverse coordinates, but has become standard through usage.
\bibitem{jenni}
The S$p$$\bar p$S program is described in the Peter Jenni'scontribution to these proceedings.
\bibitem{holmes}
The development of the Tevatron is discussed in these proceedings by Steve Holmes.
\bibitem{pepe}
The implications for the top mass from measurements at LEP were
discussed in this Symposium by Monica Pepe.
\end{thebibliography}
\end{document}